\documentclass[aps,twocolumn,10pt,showpacs,groupedaddress,superscriptaddress,floatfix,notitlepage]{revtex4-2}
\usepackage{amsmath,amsfonts,amssymb,graphics,graphicx,epsfig,color,times, braket}
\usepackage[utf8x]{inputenc}
\usepackage{color}
\usepackage{bbm, dsfont}
\usepackage{subfigure}
\usepackage{mathrsfs}
\usepackage{verbatim}
\usepackage[colorlinks=true, allcolors=blue]{hyperref}
\usepackage{graphicx}
\usepackage{dcolumn}
\usepackage{bm}
\usepackage{physics,amsmath,amssymb,enumitem,hyperref} 
\usepackage{float}

\begin{document}

\preprint{APS/123-QED}

\title{Scaling law for optimal excitation storage and superradiant release in waveguide QED systems}


\author{Wei Chen}
\email{wc2983@columbia.edu}
\affiliation{Department of Physics and Center for Theoretical Physics, National Taiwan University, Taipei, Taiwan}
\affiliation{Institute of Atomic and Molecular Sciences, Academia Sinica, Taipei 10617, Taiwan}
\affiliation{Center for Quantum Science and Engineering, National Taiwan University, Taipei 10617, Taiwan}

\author{Kuan-Ting Lin}
\email{alex968.tw@gmail.com}
\affiliation{Trapped-Ion Quantum Computing Laboratory, Hon Hai Research Institute, Taipei 11492, Taiwan}

\author{Guin-Dar Lin}
\affiliation{Department of Physics and Center for Theoretical Physics, National Taiwan University, Taipei, Taiwan}
\affiliation{Center for Quantum Science and Engineering, National Taiwan University, Taipei 10617, Taiwan}
\affiliation{Trapped-Ion Quantum Computing Laboratory, Hon Hai Research Institute, Taipei 11492, Taiwan}
\affiliation{Physics Division, National Center for Theoretical Sciences, Taipei 10617, Taiwan}

\author{H. H. Jen}
\affiliation{Institute of Atomic and Molecular Sciences, Academia Sinica, Taipei 10617, Taiwan}
\affiliation{Physics Division, National Center for Theoretical Sciences, Taipei 10617, Taiwan}
\affiliation{Department of Physics, National Central University, Taoyuan 320317, Taiwan}




\date{\today}

\begin{abstract}
Driven-dissipative quantum emitters provide a powerful platform for controllable excitation storage and release, with promising applications in quantum batteries and quantum storage. Yet, transient excitation transfer in collective many-body systems is often obscured by the intricate interplay among coherent driving, dissipation, and correlation dynamics. Here, we uncover a scalable excitation-storage mechanism in two emitter ensembles coupled to a semi-infinite waveguide. A coherently driven ensemble acts as an effective excitation reservoir, while a second ensemble positioned near a dissipative node serves as a subradiant storage medium. Surprisingly, when the driven ensemble largely exceeds the storage ensemble in size, the transfer dynamics enters a nearly correlation-free regime, allowing the driven ensemble to behave effectively as a classical excitation source. This reveals a simple scaling law for optimal excitation transfer, under which the storage ensemble approaches near-complete population inversion as the driven ensemble size increases. Building on this mechanism, we propose a three-stage storage-and-release protocol enabling fast excitation storage and controllable photon emission. Our results demonstrate how coherent and dissipative collective interactions can be jointly harnessed for quantum energy storage and programmable nonequilibrium dynamics in waveguide QED platforms.
\end{abstract}


\maketitle


\section{Introduction}

Controlled excitation storage, transfer, and release are key ingredients for the development of quantum technologies. These processes are essential for quantum memories \cite{lvovsky2009optical, zhao2009long, zhao2009millisecond}, quantum networks \cite{campaioli2024colloquium, stannigel2012driven, ramos2014quantum, pichler2015quantum}, and quantum batteries \cite{shastri2025dephasing, barra2019dissipative, farina2019charger, tabesh2020environment, xu2021enhancing, quach2022superabsorption, andolina2019extractable}, where energy or quantum information must be stored for a finite time and retrieved efficiently on demand. In this context, collective light--matter interactions provide a powerful mechanism for enhancing excitation dynamics. When multiple emitters couple to a common electromagnetic environment, such as a cavity \cite{quach2022superabsorption, wang2024cavity, chang2012cavity} or a waveguide \cite{holzinger2022control, rubies2025deterministic, mirhosseini2019cavity, sheremet2023waveguide, arcari2014near, gonzalez2024light, pichler2015quantum, lee2025controlling, jen2025photon, lalumiere2013input, lodahl2017chiral, le2021experimental, douglas2015quantum, kien2008cooperative, van2013photon, ruostekoski2016emergence}, their radiative response can become strongly modified by collective effects. In particular, superradiance enables an enhanced emission rate through constructive collective interference, while subradiance suppresses radiative decay and can protect excitations from dissipation \cite{lehmberg1970radiation, holzinger2022control, rubies2025deterministic, chen2025excitation, albrecht2019subradiant, mirhosseini2019cavity, plankensteiner2015selective, lalumiere2013input, moreno2022efficient}. The interplay between coherent excitation transfer, superradiant emission, and subradiant storage therefore offers a promising route toward fast, efficient, and controllable excitation storage-and-release protocols.

In recent years, driven-dissipative charger--battery architectures have attracted increasing attention in quantum energy-storage protocols \cite{farina2019charger, barra2019dissipative, pirmoradian2019aging, tabesh2020environment, xu2021enhancing, santos2021quantum, shastri2025dephasing}. A natural direction is to scale up the charger into a collective emitter ensemble, where cooperative light--matter interactions can enhance excitation transfer and potentially provide charging advantages \cite{peng2021lower, liu2021entanglement, yang2025optimal, liu2025entanglement}. Waveguide quantum electrodynamics (QED) systems are particularly suitable for this purpose because photon-mediated interactions are long-ranged and can be engineered through the emitter positions and boundary conditions. In a semi-infinite waveguide, the coherent and dissipative parts of the dipole--dipole interaction (DDI) can be spatially controlled, allowing radiatively active emitters to serve as chargers while subradiant emitters store excitations \cite{lin2019scalable,lin2022deterministic, lin2025single}. This makes the platform a promising candidate for realizing collective charging, protected storage, and controlled release in a driven-dissipative setting.

However, identifying the optimal excitation-transfer condition becomes increasingly challenging as the system size grows. In the early-time dynamics, coherent driving, dissipation, and inter-ensemble coupling act simultaneously, while the transfer efficiency depends sensitively on the buildup of inter-ensemble correlations \cite{shastri2025dephasing,farina2019charger,tabesh2020environment,xu2021enhancing,mayo2022collective,canzio2025single,andolina2019extractable,lehmberg1970radiation, jen2022quantum, handayana2025suppression}. These correlations generally invalidate a simple mean-field description, making an analytical characterization of the optimal transfer condition difficult in the strong-driving regime. Despite this challenge, we show that a semi-infinite-waveguide platform reveals an unexpectedly simple route to optimal excitation transfer. We consider two emitter ensembles coupled through the waveguide: a coherently driven ensemble that supplies excitations, and a storage ensemble positioned near a dissipative node, where it becomes subradiant and protected from direct radiative loss. By analyzing the inter-ensemble correlation, we uncover a nearly uncorrelated transfer regime that emerges when the driven ensemble is much larger than the storage ensemble. In this regime, the driven ensemble acts as an effective classical excitation source during the early-time dynamics, making the otherwise complex many-body transfer process amenable to a mean-field description. This leads to a simple scaling law for the optimal transfer condition, under which the storage ensemble approaches full inversion as the driven ensemble is scaled up. Building on this mechanism, we further develop a storage-and-release protocol in which excitations are coherently transferred into the subradiant storage ensemble, protected by detuning the driven ensemble, and released on demand by restoring resonance, with the output enhanced through the superradiant decay channel of the driven ensemble.

The paper is organized as follows. In Sec.~\ref{theo_model}, we introduce the semi-infinite waveguide setup and derive the effective master equation for two collective emitter ensembles. In Sec.~\ref{early_time}, we study the early-time excitation-transfer dynamics and show that the storage ensemble can approach full inversion under an optimized driving condition. In Sec.~\ref{optimal_condition}, we analyze the inter-ensemble correlation and identify the nearly uncorrelated transfer regime that leads to a simple scaling law for the optimal ratio between the waveguide-mediated coupling and the driving strength. In Sec.~\ref{storage_and_release}, we use this mechanism to construct a storage-and-release protocol and characterize the resulting superradiant emission during the release stage. In Sec.~\ref{exp_realization}, we discuss experimental requirements and possible imperfections, including operation timescales and additional decoherence channels. Finally, Sec.~\ref{conclusion} summarizes our results and discusses their implications for excitation storage and controlled emission in waveguide QED platforms.

\section{Theoretical model}
\label{theo_model}
\begin{figure}[t]
    \centering
    \includegraphics[width=1\linewidth]{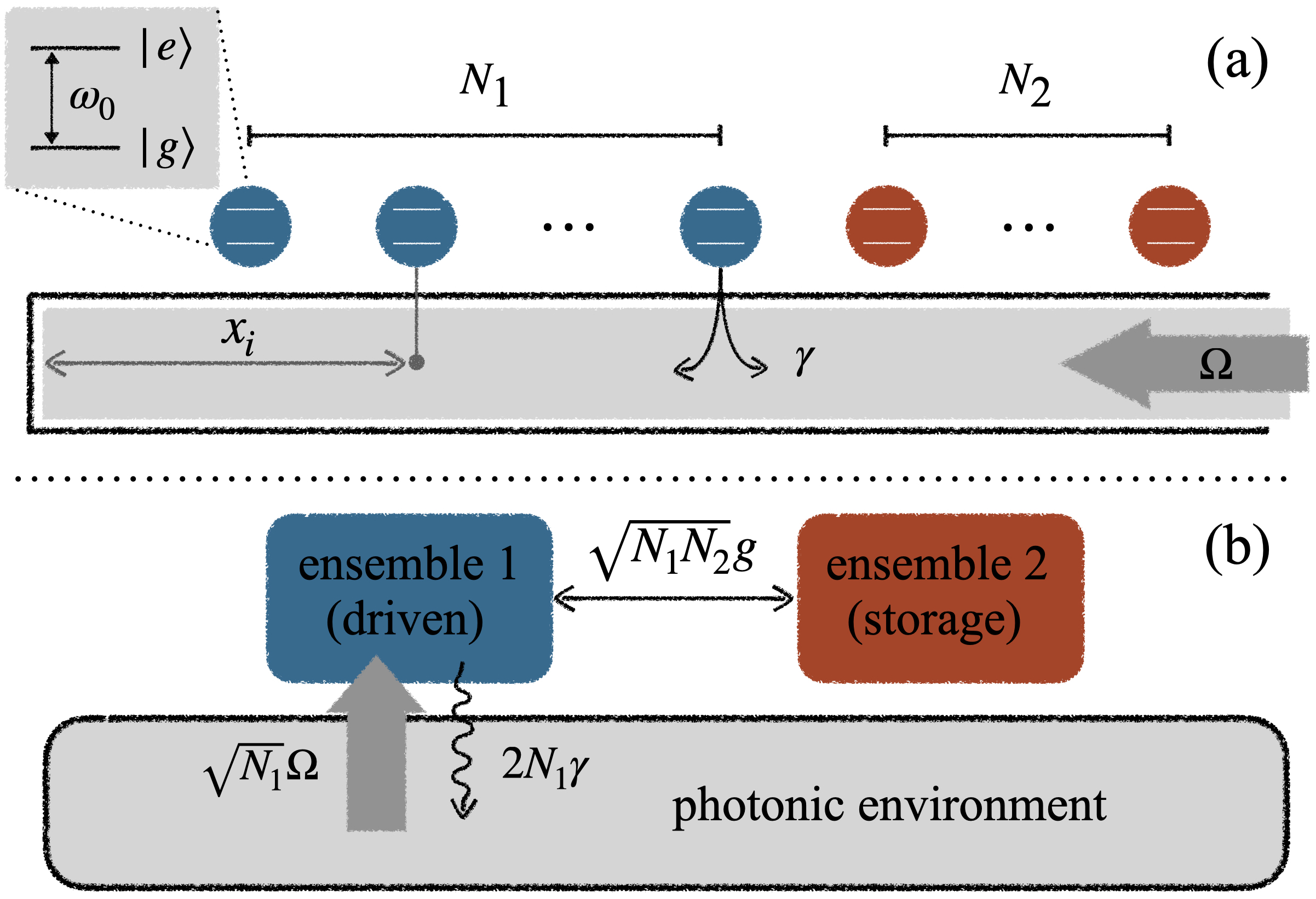}
    \caption{\textbf{Schematic of the system setup}. (a) Two emitter arrays, driven and storage ensemble, consisting of $N_1$ and $N_2$ emitters, interacting via an infinite-range DDI mediated by a 1D semi-infinite waveguide (with individual emitter decay rate of $\gamma$). Each emitter is a two-level system with ground state $\ket{g}$, excited state $\ket{e}$, and transition frequency $\omega_0$. The distance of the $i$-th emitter from the closed end of the waveguide is $x_i$. Additionally, the system is driven by a coherent waveguide field from the open end with driving strength of $\Omega$. (b) In the case of Eq.~(\ref{special_config}), the system turns into two coupled collective spin, ensemble 1 (driven) and 2 (storage). Driven ensemble is driven-dissipative with driving strength of $\sqrt{N_1}\Omega$ and dissipation rate of $2N_1\gamma$.}
    \label{config}
\end{figure}

We consider two spatially separated quantum emitter ensembles--the driven and storage ensembles--containing $N_1$ and $N_2$ two-level emitters, respectively. These ensembles interact with a one-dimensional waveguide featuring an antinode mirror at $x = 0$, enabling effective inter-ensemble coupling via DDI mediated by continuous waveguide modes, as illustrated in Fig.~\ref{config}. All two-level quantum emitters are identical, having an excited state $\ket{e}$ and a ground state $\ket{g}$ separated by transition frequency $\omega_0$ (wavelength $\lambda_0$). A coherent field with frequency $\omega_d$ and driving strength $\Omega$ is incident from the open end to pump the ensembles. The system dynamics is governed by the following Markovian master equation for the system's density matrix $\hat{\rho}$ (with $\hbar = 1$) \cite{lin2019scalable, lehmberg1970radiation}:
\begin{align}
    \dot{\hat{\rho}} = 
    & i\sum_i \delta \left[\hat{\sigma}^\dagger _i\hat{\sigma}_i, \hat{\rho} \right] + i \sum_{i} \Omega \cos(k_dx_i)\left[\hat{\sigma}^\dagger_i + \hat{\sigma}_i, \hat{\rho} \right] \notag \\
    & - i \sum_{ij}g_{ij}\left[\hat{\sigma}^\dagger_i \hat{\sigma}_j, \hat{\rho} \right] +  \sum_{ij} 2\gamma_{ij} \mathcal{L}_{ij}[\hat{\rho}],
    \label{ME_1}
\end{align}
where $\delta = \omega_d - \omega_0$ is the atom-field detuning and $k_d = \omega_d/c$ is the wave number associated with the waveguide speed of light $c$. The lowering operator for the $i$-th emitter at position $x_i$ is defined as $\hat{\sigma}_i = \ket{g_i}\bra{e_i} = (\hat{\sigma}_i^\dagger)^\dagger$. Furthermore, the DDI-induced dissipative and coherent coupling rates are given by:
\begin{equation}
\begin{cases}
    \gamma_{ij} = \frac{\gamma}{2} \left[ \cos k_0(x_i + x_j) + \cos k_0 \left|x_i-x_j \right|  \right], \\
    g_{ij} = \frac{\gamma}{2} \left[ \sin k_0(x_i + x_j) + \sin k_0 \left|x_i-x_j \right|  \right], 
    \label{DDI_strength}
\end{cases}
\end{equation}
in which  $\gamma$ is the individual decay rate of the emitters into the waveguide, $k_0 = \omega_0/c$ is the wavevector, and $\mathcal{L}_{ij} = \hat{\sigma}_i\hat{\rho}\hat{\sigma}_j^\dagger - \{\hat{\sigma}_i^\dagger \hat{\sigma}_j, \hat{\rho} \}/2$ is the superoperator describing the collective dissipation. The effect of the mirror can be thought of as placing image emitters on the other side of the mirror, thus the DDI strength between the $i$-th and $j$-th emitter depends on both the relative distance $\big| x_i - x_j \big|$ and the image one $(x_i + x_j)$.

In this work, we investigate the inter-ensemble excitation dynamics and excitation storage under a resonant coherent driving field ($\omega_d = \omega_0$) is a specific configuration, where the spatial arrangement of the two-level emitters is chosen as  
\begin{equation}
\begin{cases}
    x_i = (i+1/2)\lambda_0, \text{ for } i \in \{1\}, \\
    x_i = (i+3/4)\lambda_0, \text{ for } i \in \{2\},
\end{cases}
\label{special_config}
\end{equation}
where $i = 1,2, ..., N_1+N_2$, and $\{1\} = \{1, 2, ..., N_1\}$ and  $\{2\} = \{N_1+1, N_1+2, ..., N_1+N_2\}$ are the driven and storage ensemble, respectively. Specifically, the emitters belonging to the driven and storage ensembles are positioned exactly at the antinodes and nodes of the resonant waveguide field, respectively. Note that the driven ensemble is placed closer to the closed end of the waveguide than the storage ensemble. By introducing the collective operators $\hat{S}_{1(2)} = \sum_{i \in \{1(2)\}} \hat{\sigma}_i/\sqrt{N_{1(2)}}$, Eq.~(\ref{ME_1}) can be recast into
\begin{align}
    \dot{\hat{\rho}} = 
    & -i \Omega \sqrt{N_1} \left[\hat{S}^\dagger_1 + \hat{S}_1, \hat{\rho} \right] \notag \\
    & - i g \sqrt{N_1N_2}\left[\hat{S}^\dagger_1 \hat{S}_2 + \hat{S}^\dagger_2 \hat{S}_1, \hat{\rho} \right] + 2N_1\gamma \mathcal{L}_{1}[\hat{\rho}],
    \label{ME_2}
\end{align}
where $\mathcal{L}_1[\hat{\rho}] =  \hat{S}_1 \hat{\rho} \hat{S}_1^\dagger - \frac{1}{2} \left\{ \hat{S}^\dagger_1\hat{S}_1, \hat{\rho} \right\}$ denotes the collective dissipation. Here, to avoid any potential confusion, we explicitly denote the coherent exchange coupling rate as $g=\gamma$. As governed by Eq.~(\ref{ME_2}),  this effective model features a driven-dissipative collective spin (driven ensemble) that is pumped by a coherent field with an enhanced driving strength $\sqrt{N_1}\Omega$ and undergoes collective dissipation at a rate of $2N_1\gamma$. Furthermore, this spin is coherently coupled to a second, non-dissipative collective spin (storage ensemble), enabling excitation exchange at a collectively enhanced rate of $\sqrt{N_1N_2}g$ [see Fig.~\ref{config}(b)].

\section{Excitation dynamics}
\label{early_time}
\begin{figure}[t]
    \centering
    \includegraphics[width=1\linewidth]{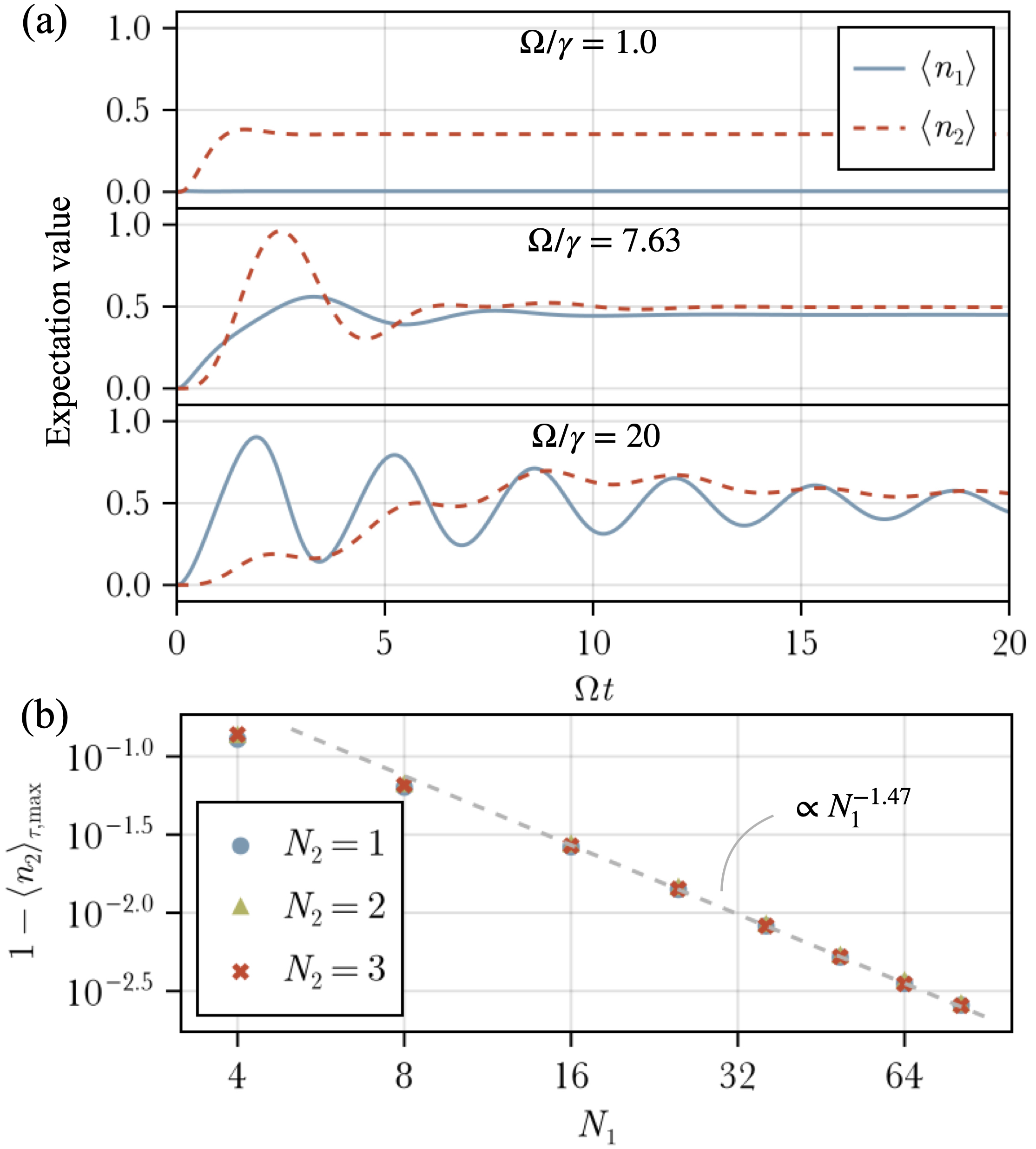}
    \caption{\textbf{Time evolution and scaling law of the excitation number in storage ensemble.}. 
    (a) Time evolution of the excitation number $\langle n_2\rangle$ for a system with
    $(N_1,N_2)=(12,1)$ at $\Omega/\gamma = 1.0$, $7.63$, and $20$, shown from top to bottom. (b) Deviation of the maximum excitation number from unity,
    $1-\langle n_2\rangle_{\tau,\max}$, as a function of $N_1$ for different
    system sizes $(N_1,N_2)$. We simulate $N_2=1,2,$ and $3$, and find that the
    results converge in the limit $N_1\gg N_2$.}
    \label{TE_demo}
\end{figure}
In this section, we aim to investigate the dynamics of the emitter ensembles under various driving strengths. As shown in Fig.~\ref{TE_demo}, we monitor the time evolution of the mean excitation number, defined as $\langle n_{1(2)} \rangle = (\langle \hat{S}^z_{1(2)}\rangle + 1)/2$, where $\hat{S}^z_{1(2)} = \sum_{i\in\{1(2)\}}\hat{\sigma}^z_i/N_{1(2)}$ and $\hat{\sigma}^z_i = \ket{e_i}\bra{e_i} - \ket{g_i}\bra{g_i}$. Throughout the following sections, the system's dynamics are obtained by numerically solving the master equation in Eq.~(\ref{ME_2}).

To examine the effects of different driving strengths on excitation transfer between the driven and storage ensembles, we consider a representative case where $(N_1, N_2) = (12, 1)$. According to the effective model in Eq.~(\ref{ME_2}), the relative strength between the inter-ensemble coupling ($\sqrt{N_1N_2}g$) and the coherent drive ($\sqrt{N_1}\Omega$) is directly characterized by the ratio $\Omega/\sqrt{N_2}\gamma$. Since $N_2 = 1$, this ratio is exactly $\Omega/\gamma$. Consequently, in the drive-dominated regime where $\Omega/\gamma = 20$, the inter-ensemble interaction is too weak to mediate efficient transfer. As shown in the bottom panel of Fig.~\ref{TE_demo}(a), this causes $\langle n_1 \rangle$ to undergo several rapid Rabi oscillations before the excitation is significantly transferred to $\langle n_2 \rangle$.

Conversely, if the driving strength is excessively weakened, such as $\Omega/\gamma = 1.0$, the rapid collective relaxation process (with a rate of $N_1\gamma = 12\gamma > \sqrt{N_1}\Omega \approx 3.46\gamma$) overwhelms the effective drive. As displayed in the top panel of Fig.~\ref{TE_demo}(a), $\langle n_1 \rangle$ is barely populated. Although a noticeable population still builds up in $\langle n_2 \rangle$, its maximum value remains limited, indicating that a stronger drive is required to feed sufficient excitations into the driven ensemble.

As a result, to effectively populate the storage ensemble, the driving strength must be carefully optimized to balance these coherent and dissipative timescales. Indeed, as depicted in the middle panel of Fig.~\ref{TE_demo}(a), tuning the ratio to the optimal value of $\Omega/\gamma \approx 7.63$ allows the stored excitation $\langle n_2 \rangle$ to transiently approach unity.

Beyond optimizing the driving strength, we can further engineer the collective exchange and dissipation rate by varying the number of emitters in each ensemble. To investigate this size effect, we define $\tau$ as the time at which $\langle n_2 (t)\rangle$ reaches its global maximum during the time evolution. By scanning different values of $\Omega/\gamma$, we identify the optimal ratio $(\Omega/\gamma)_{\mathrm{max}}$ that maximizes $\langle n_2\rangle_{\tau} = \langle n_2 (t=\tau)\rangle$. Fig.~\ref{TE_demo}(b) shows that the optimized maximal excitation number $\langle n_2\rangle_{\tau,\max}$ approaches unity as $N_1$ increases. Numerically, we find the scaling relation $1-\langle n_2\rangle_{\tau,\max} \propto N_1^{-1.47}$ in the highly asymmetric limit $N_1\gg N_2$, demonstrating that the size of the driven ensemble governs the ultimate transfer efficiency. This scaling behavior indicates that emitters in storage ensemble can be excited close to the fully inverted state indirectly through the coherent interaction with driven ensemble.

\begin{figure}[t]
    \centering
    \includegraphics[width=1\linewidth]{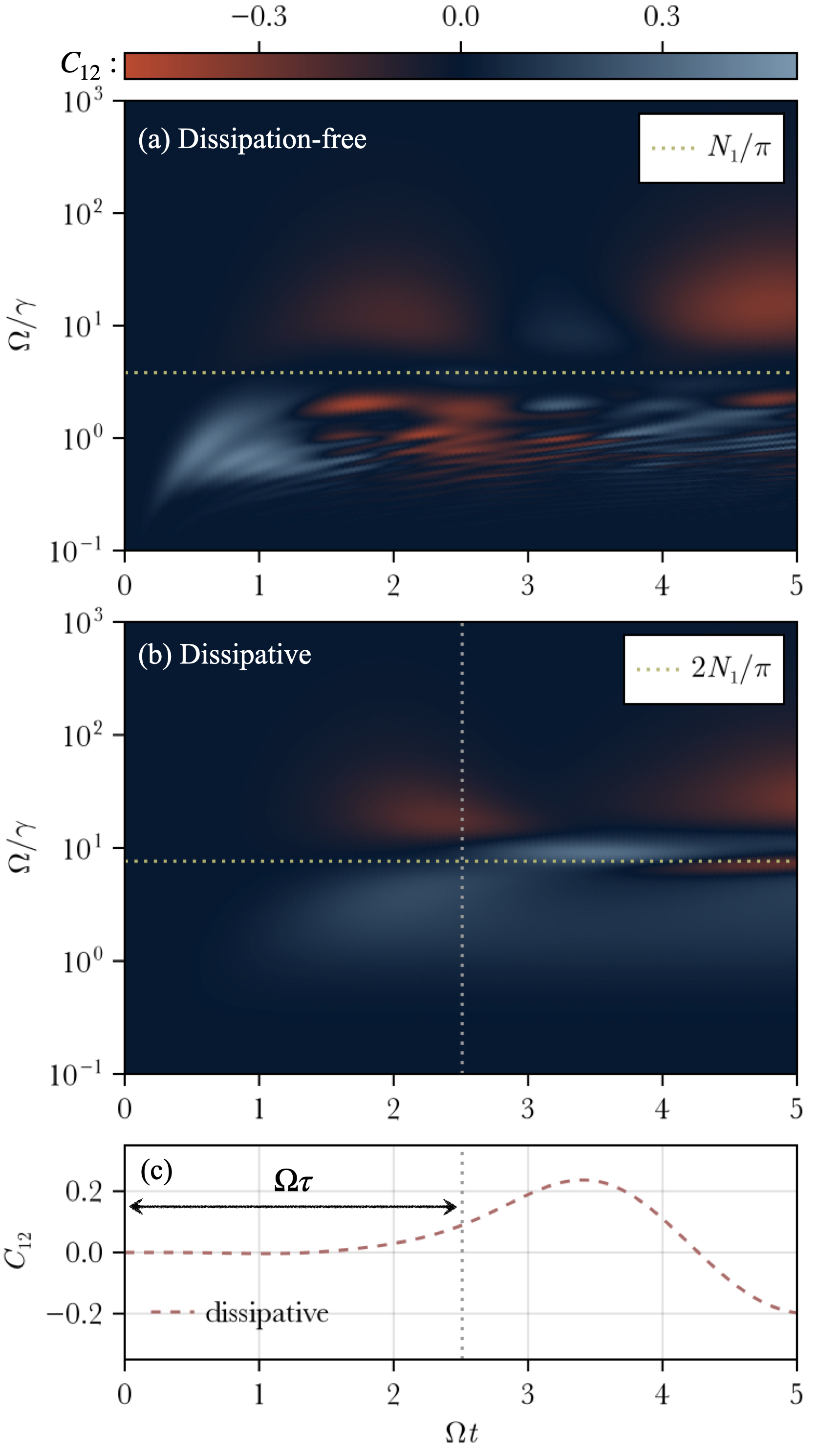}
    \caption{\textbf{Time evolution of the inter-ensemble correlator in different coupling regimes.}
    We calculate the connected correlator $C_{12}$ from the density matrix obtained from Eq.~\eqref{ME_2}. Panels (a) exclude the dissipative contribution, whereas panel (b) shows the dynamics including the effect of the dissipator. The system size is $(N_1,N_2)=(12,1)$ in both panels. The yellow dotted line in (a) indicates the transition point separating different coupling regime and the one in (b) marks the optimal ratio $(\Omega/\gamma)_\text{max}$. By comparing (a) and (b), we further show that the dissipator induces inter-ensemble mixing and leads to the accumulation of correlations during the time evolution. (c) Time evolution of \(C_{12}\) along the dotted lines shown in panels (b). The vertical gray dotted lines in panels (b) and (c) indicate the optimal driving duration for \(\Omega/\gamma = (\Omega/\gamma)_{\mathrm{max}}\).
    }
    \label{correlator_TE}
\end{figure}

\section{Optimal excitation transfer condition}
\label{optimal_condition}

In Sec.~\ref{early_time}, we numerically demonstrated that the excitation transfer to the storage ensemble can be optimized to approach near-unity by carefully tuning the ratio $\Omega/\gamma$. This numerical observation motivates us to seek a rigorous analytical understanding of this optimal driving condition and its underlying physical mechanism. Generally, the DDI dynamically generates strong correlations between the two ensembles. To quantify this, we examine the connected inter-ensemble correlator, defined as \cite{kubo1962generalized, schack1990moment}:
\begin{equation}
    C_{12} =
    \langle \hat{S}^{\dagger}_1 \hat{S}_2 \rangle
    -
    \langle \hat{S}^{\dagger}_1 \rangle
    \langle \hat{S}_2 \rangle .
\end{equation}

This second-order cumulant measures the genuine quantum fluctuations beyond the simple product of their mean fields. Crucially, we find that a nearly correlation-free transfer regime can emerge in the highly asymmetric limit ($N_1 \gg N_2$), provided the driving strength is properly tuned, as shown in Fig.~\ref{correlator_TE}. In this macroscopic limit, when the dynamics of the overwhelmingly large driven ensemble are successfully dominated by the coherent drive, it effectively acts as a macroscopic, classical excitation source for the storage ensemble, much like a classical laser field driving an atom. By targeting this specific parameter space where quantum correlations are persistently suppressed ($C_{12} \approx 0$), we can rigorously validate the use of a mean-field approximation to analytically derive the optimal early-time transfer condition.

\subsection{Dissipation-free case}

To examine the role of the collective exchange rate in the storage efficiency, we first isolate the coherent dynamics by omitting the dissipation term in Eq.~(\ref{ME_2}). As shown in Fig.~\ref{correlator_TE}(a), the inter-ensemble correlation $C_{12}$ is  suppressed along a characteristic ratio of $\Omega/\gamma$ throughout the time evolution, as indicated by the dotted line. By targeting this specific correlation-free regime, we can rigorously apply the mean-field approximation \cite{kubo1962generalized} to factorize the expectation values as $\langle \hat{O}_1\hat{O}_2\rangle \approx \langle \hat{O}_1 \rangle \langle \hat{O}_2 \rangle$, where $\hat{O}_i$ represents the collective operator $\hat{S}_i$ or $\hat{S}^z_i$ for $i = 1,2$. The simplified system dynamics are then described by:
\begin{equation}
    \begin{cases}
        \langle\dot{\hat{S}}^z_1\rangle
        =
        \frac{-4\Omega}{\sqrt{N_1}}
        \operatorname{Im}\langle \hat{S}_1\rangle
        +
        4g\sqrt{\frac{N_2}{N_1}}
        \operatorname{Im}
        \left[
        \langle\hat{S}_1\rangle^*
        \langle\hat{S}_2\rangle
        \right],
        \\[6pt]
        \langle\dot{\hat{S}}_1\rangle
        =
        i\left[
        \sqrt{N_1}\Omega
        \langle\hat{S}^z_1\rangle
        +
        g\sqrt{N_1N_2}
        \langle\hat{S}_2\rangle
        \langle\hat{S}^z_1\rangle
        \right],
        \\[6pt]
        \langle\dot{\hat{S}}^z_2\rangle
        =
        4g\sqrt{\frac{N_1}{N_2}}
        \operatorname{Im}
        \left[
        \langle\hat{S}_1\rangle^*
        \langle\hat{S}_2\rangle
        \right],
        \\[6pt]
        \langle\dot{\hat{S}}_2\rangle
        =
        ig\sqrt{N_1N_2}
        \langle \hat{S}_1\rangle
        \langle \hat{S}^z_2\rangle .
    \end{cases}
    \label{mf_eom}
\end{equation}

To analytically solve the equations, we further normalize the collective atomic coherence by dividing $\langle \hat{S}_1 \rangle$ by $N_1$ in Eq.~(\ref{mf_eom}). Under this normalization, the effective inter-ensemble exchange rate governing the dynamics of the driven ensemble explicitly scales as $g\sqrt{N_2/N_1}$. Consequently, in the highly asymmetric macroscopic limit $N_1 \gg N_2$, the back-action from the storage ensemble onto the driven ensemble becomes strictly negligible. This mathematically demonstrates that the evolution of the driven ensemble is overwhelmingly dominated by the coherent drive $\Omega$, allowing it to dynamically decouple from the inter-ensemble interaction. Using the initial conditions $\langle\hat{S}^z_1\rangle_0=\langle\hat{S}^z_2\rangle_0=-1$ and $\langle\hat{S}_1\rangle_0=\langle\hat{S}_2\rangle_0=0$, we directly integrate the simplified equations to obtain:
\begin{equation}
    \begin{cases}
        \langle \hat{S}^z_1\rangle
        =
        -\cos(2\Omega t),
        \\[6pt]
        \langle \hat{S}_1 \rangle
        =
        -\dfrac{i}{2}\sqrt{N_1}\sin(2\Omega t),
        \\[6pt]
        \langle \hat{S}^z_2\rangle
        =
        -\cos\!\left[
        \dfrac{gN_1}{\Omega}
        \left(1-\cos(2\Omega t)\right)
        \right],
        \\[6pt]
        \langle \hat{S}_2\rangle
        =
        -\sqrt{N_2}\sin\!\left[
        \dfrac{gN_1}{\Omega}
        \left(1-\cos(2\Omega t)\right)
        \right],
    \end{cases}
    \label{mf_solution}
\end{equation}
from which the optimized transfer condition is determined by solving $\langle \hat{S}^z_2\rangle=1$, yielding
\begin{equation}
    \Omega t
    =
    \frac{1}{2}
    \arccos\!\left(
    1-\frac{2\pi\Omega}{gN_1}
    \right).
    \label{t_equation}
\end{equation}

For a physically valid (real-valued) time t to exist, this imposes a strict upper bound on the coupling ratio: $\Omega/g \le N_1/\pi$. This optimized condition is clearly illustrated by the numerical results of the correlator $C_{12}$ for the $(N_{1},N_{2})=(12,1)$ case in Fig.~\ref{correlator_TE}(a). Within the strong driving regime where $\Omega/g = \Omega/\gamma > 12/\pi \approx 3.82$, the inter-ensemble correlation is not universally suppressed. Moreover, our analysis based on Eq.~(\ref{t_equation}) reveals that the exchange coupling here is simply too weak to achieve complete excitation transfer. Conversely, as the system moves into the strong coupling regime, $\Omega/g < 3.82$, the back-action dominates the excitation transfer, leading to the buildup of strong correlations. Therefore, the optimal, correlation-free excitation transfer exclusively emerges exactly at the boundary separating these two regimes ($\Omega/g \approx 3.82$), as indicated by the dotted line. At this unique theoretical sweet spot, the system remarkably decouples to ensure full transfer while strictly validating our mean-field assumption.

\subsection{Dissipative dynamics and modified transfer condition}

We now turn to a more realistic scenario where the collective decay process must be taken into account. This collective effect can further mix the two ensembles, as displayed in Fig.~\ref{correlator_TE}(b). We observe that the two ensembles become strongly correlated, and the strictly correlation-free regime vanishes entirely. Consequently, the mean-field approximation breaks down, rendering Eq.~(\ref{mf_eom}) invalid and making a simple analytical solution unattainable.

\begin{figure}[t]
    \centering
    \includegraphics[width=1\linewidth]{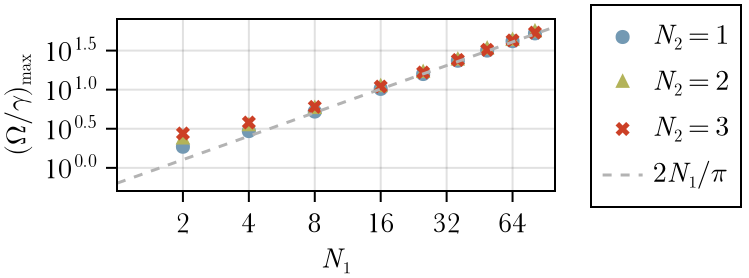}
    \caption{\textbf{Scaling law of the optimal ratio $(\Omega/\gamma)_{\mathrm{max}}$.}
    By numerically simulating the system dynamics according to Eq.~\eqref{ME_2}, we extract the optimal ratio between the coupling strength $\gamma$ and the driving strength $\Omega$. The numerical results are compared with the scaling relation in Eq.~\eqref{optimal_ratio}, showing good agreement in the regime $N_1\gg N_2$.
    }
    \label{scaling_law}
\end{figure}

However, as demonstrated in Fig.~\ref{TE_demo}(a), the population can be efficiently transferred to the storage ensemble during the early stages of the system's dynamics by employing an optimized driving strength. Consequently, our primary focus lies in this early-time optimal transient excitation transfer. For evolution times $t \le \tau$, where $\tau$ denotes the time at which the maximal excitation transfer is achieved, we observe that the correlator $C_{12}$ remains vanishingly small. This implies that the excitation transfer process completes before collective dissipation can induce substantial inter-ensemble correlations; the system remains barely correlated, as explicitly displayed in Fig.~\ref{correlator_TE}(b) and \ref{correlator_TE}(c). In this transient regime, the mean-field approximation remains robust, justifying the applicability of the analytic framework from Eq.~(\ref{mf_eom})--(\ref{t_equation}). As a result, the optimal driving condition elegantly obeys the $N_1$ scaling law. To precisely quantify this behavior, we calculate the optimal driving strength for maximal excitation transfer across various emitter numbers $(N_1, N_2)$, as shown in Fig.~(\ref{scaling_law}). By numerically fitting these results, we find that the optimal driving condition follows
\begin{equation}
    \left( \frac{\Omega}{\gamma} \right)_{\mathrm{max}}
    \approx
    \frac{2N_1}{\pi}.
    \label{optimal_ratio}
\end{equation}
This indicates that a nearly twofold stronger coherent drive (compared to the dissipation-free limit $N_{1}/\pi$) is required to exceed the collective decay and achieve efficient excitation transfer. Here, we emphasize that even at this numerically optimized driving strength (e.g., $\Omega/\gamma \approx \pi/24 \approx 7.63$ for $N_{1}=12$), significant inter-ensemble correlations persist, as indicated by the dotted line in Fig.~\ref{correlator_TE}(b). This means that the optimal transfer in the presence of dissipation is no longer a purely macroscopic, correlation-free process. Instead, it represents a delicate trade-off. The drive must be strong enough to overcome the excitation loss, yet this stronger drive inevitably reintroduces the inter-ensemble back-action, leading to a correlation buildup that inherently limits the ultimate storage efficiency.

\section{Excitation storage and release}
\label{storage_and_release}

\begin{figure}[t]
    \centering
    \includegraphics[width=1\linewidth]{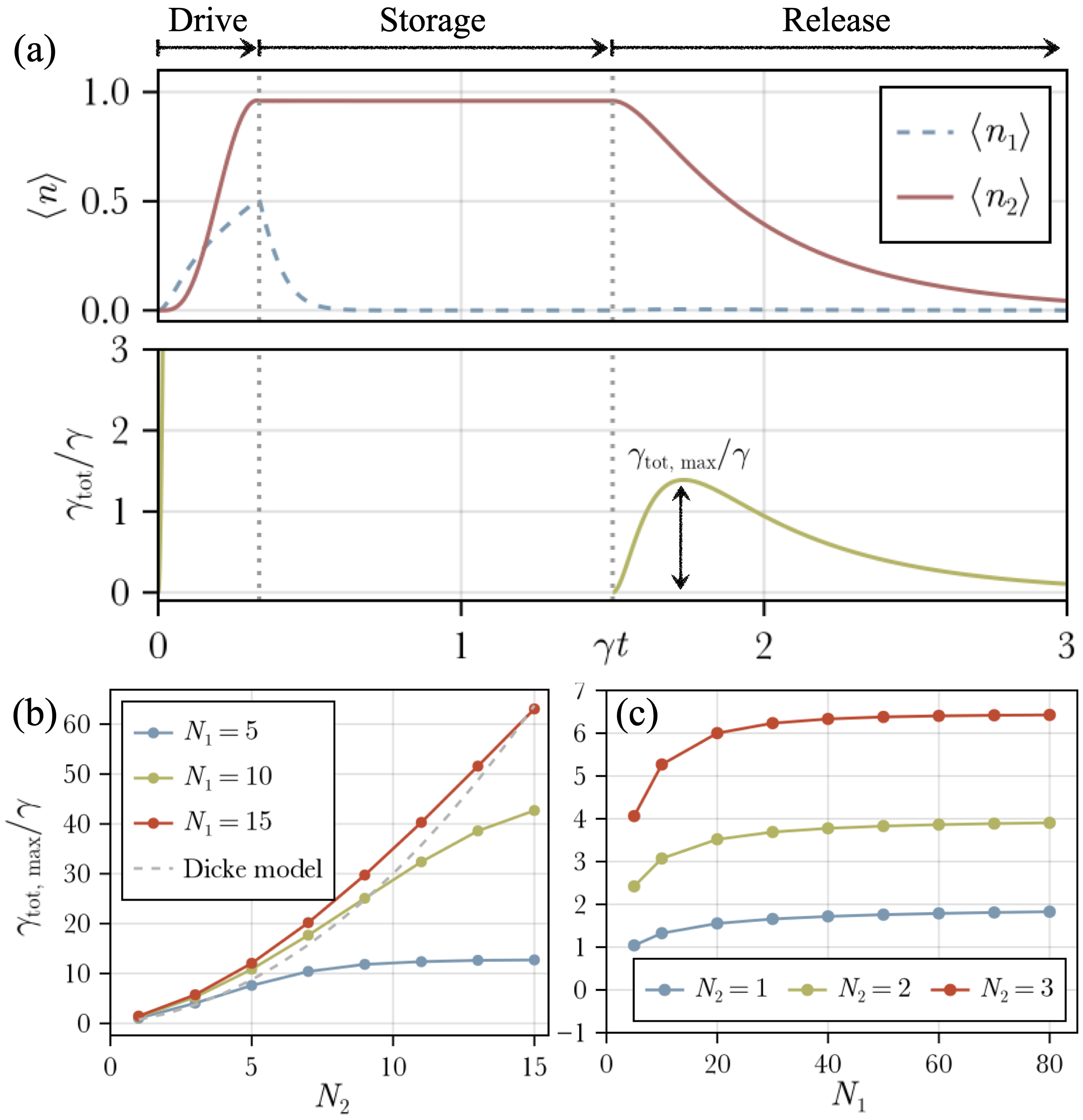}
    \caption{
    \textbf{Excitation storage and release process.}
    (a) Time evolution of the excitation numbers and the emission rate, with system size $(N_1, N_2) = (12, 1)$. During the driving stage, the system is driven through the waveguide at the optimal ratio of $\Omega/\gamma$. At the end of this stage, the drive is turned off and the emitters in driven ensemble are detuned far from the original resonance frequency, with $\Delta/\gamma=500$, thereby storing the excitation in storage ensemble. After the storage stage, the emitters in driven ensemble are tuned back into resonance, allowing the excitation stored in storage ensemble to be released. 
    (b,c) Peak emission rate during the release process for different system sizes.
    }
    \label{storage_release}
\end{figure}

Building upon the optimal scaling law uncovered above, we propose a programmable excitation storage-and-release protocol, inspired by Holzinger et al. \cite{holzinger2022control}, that capitalizes on our early-time transfer mechanism. The protocol comprises three distinct operational stages—drive, storage, and release—as illustrated by the time evolution of the ensemble populations and the total emission rate $\gamma_\text{tot}=2N_1\gamma\langle \hat{S}_1^{\dagger}\hat{S}_1 \rangle$ in Fig.~\ref{storage_release}(a).

During the drive stage, we illuminate the driven ensemble using a photon pulse with an optimal driving strength, exploiting the collective DDI to route the excitation into the storage ensemble. Once optimal population inversion is achieved, we initiate the storage stage by quenching the inter-ensemble dynamics: the drive is abruptly switched off, and a large frequency detuning is introduced between the two ensembles. Crucially, the storage ensemble is positioned at the nodes of the resonant modes such that its emission destructively interferes with its own mirror image, isolating the stored excitation from the waveguide continuum. Finally, to trigger the release stage, we tune the driven ensemble back into resonance with the storage ensemble. The stored excitation is coherently transferred back to the driven ensemble, which then rapidly dissipates the energy into the waveguide via its macroscopic, superradiant decay channel.

The efficiency of this retrieval process is characterized by the peak emission rate, $\gamma_\text{tot,max}$ [indicated by an arrow in Fig.~\ref{storage_release}(a)], which exhibits a strong dependence on the relative ensemble sizes, as displayed in Fig.~\ref{storage_release}(b). To benchmark this performance, we compare $\gamma_\text{tot,max}$ against the standard Dicke superradiance limit for an isolated collective spin of size $N_2$, which yields a peak rate of $\gamma_D=N_2(N_2/2+1)\gamma/2$. We observe that the emission rate is significantly enhanced beyond this Dicke limit when $N_1>N_2$, suggesting that the macroscopic driven ensemble acts as a superradiant amplifier for the released photon. Furthermore, in the highly asymmetric limit $N_1 \gg N_2$, this peak emission rate asymptotically saturates across different storage sizes, as shown in Fig.~\ref{storage_release}(c). For instance, in the fundamental limit of single-excitation storage ($N_2=1$), an exact analytical treatment (detailed in Appendix \ref{N_2=1}) reveals that the normalized peak emission rate strictly converges to $\gamma_\text{tot,max}/\gamma \rightarrow 2$ as $N_1 \rightarrow \infty$.

\section{Experimental realization}
\label{exp_realization}
\subsection{Experimental parameters and operations}

\begin{figure}[t]
    \centering
    \includegraphics[width=1\linewidth]{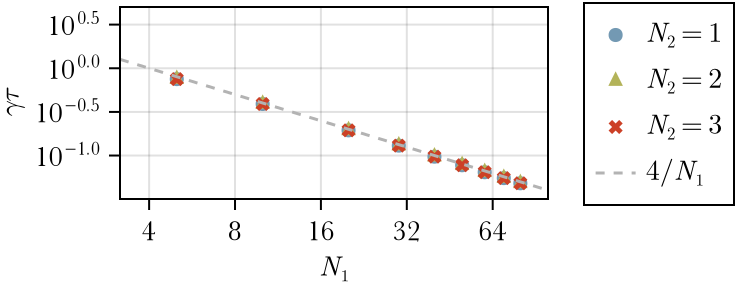}
    \caption{
    \textbf{Optimal driving duration $\tau$.}
    For the case of $N_1 \gg N_2$ the optimal driving duration scales $\sim 4/N_1$ (dashed gray line) with the emitter number in driven ensemble.
    }
    \label{tau_scaling}
\end{figure}

To physically realize the proposed storage-and-release
protocol, precise control over the driving strength and
pulse duration is required. Superconducting quantum
circuit architectures provide an ideal platform for such
implementations, as the concept of enhancing emitter interactions via a mirror in a waveguide has already been
demonstrated in these systems \cite{wen2019large}. To optimize the excitation transfer to the storage ensemble, we first determine the optimal driving duration $\tau$ against the driven ensemble size $N_1$ as displayed in Fig.~\ref{tau_scaling} for various storage
ensemble sizes ($N_2 = 1, 2, 3$). By numerical fitting, we
find that in the highly asymmetric limit ($N_1 \gg N_2$) , the duration satisfies a simple scaling relation of $\tau \approx 4/(N_1\gamma)$.

Rather than assuming hypothetical coupling strengths, our scheme can be rigorously grounded in accessible experimental parameters. Taking a recent experiment as a representative case with $(N_1, N_2) = (12, 1)$. and an individual relaxation rate of $\gamma = 2\pi \times 0.5$ MHz \cite{lu2021propagating}, we obtain an optimal pulse duration of $\tau = 106$ ns. Notably, our definition of the relaxation rate \cite{lin2019scalable} is half of the population decay rate conventionally reported in Ref. \cite{lu2021propagating}; therefore, we adopt $\gamma = 2\pi \times 0.5$ MHz in our setup. Correspondingly, to satisfy the optimal excitation transfer condition $(\Omega/\gamma)_\text{max} \approx 7.63$, one can utilize an arbitrary waveform generator to apply a square pulse with a driving strength of $\Omega \approx 7.6\gamma \approx 2\pi \times 3.8$ MHz from the open end of the waveguide.

Following the optical pumping process, one can apply
external magnetic fluxes to rapidly tune the resonance
frequencies of each emitter in the driven ensemble, largely
detuning them from the storage ensemble within $10$ ns \cite{blais2021circuit, huang2025tunable}. As a result, the excitation can be efficiently stored in the storage ensemble due to destructive interference with its mirror image. Finally, to release the excitation, one can rapidly tune the driven ensemble back into resonance with the storage ensemble. The stored excitation can then be rapidly dissipated from the driven ensemble via superradiant relaxation. 

Within this operational regime, the influence of the
non-radiative decay rate $\gamma_\text{nr}$ and the pure dephasing rate $\gamma_\phi$ can be safely neglected (see Appendix~\ref{nr_phi_deacy} for a detailed analysis), provided that the waveguide-mediated coupling dominates the system dynamics, i.e., $\gamma \gg \gamma_\text{nr},\gamma_\phi$. This parameter regime is readily accessible using state-of-the-art superconducting architectures.

\subsection{Robustness against pure dephasing}

Here, we highlight a crucial advantage of our protocol: its inherent robustness against pure dephasing processes compared to conventional subradiant-state storage schemes. In many subradiant or dark-state protocols, excitation protection relies heavily on maintaining a coherent, antisymmetric superposition between spatially separated emitters. Such spatially distributed entanglement makes the stored excitations highly susceptible to dephasing-induced loss of relative phase coherence \cite{holzinger2022control, chen2025excitation}. In contrast, our protocol is fundamentally based on the destructive interference between the storage ensemble and its own mirror image to achieve subradiant-state protection. Because this interference is rigidly imposed
by the boundary conditions of the semi-infinite waveguide, local pure dephasing does not easily destroy the relative phase between an emitter and its mirror image. As a result, pure dephasing affects the protocol much less severely than non-radiative decay, becoming detrimental only when the dephasing rate is sufficiently large to disrupt the early-time coherent transfer or the final superradiant release dynamics.

\section{Conclusion}
\label{conclusion}
In conclusion, we have investigated the underlying mechanism for scalable excitation transfer and storage in a semi-infinite waveguide QED system consisting of two spatially separated, collectively coupled emitter ensembles—the driven and the storage. By positioning the emitters of the storage ensemble exactly at the nodes of the resonant mode, their radiative decay is strongly suppressed through destructive interference with their mirror images. Simultaneously, the storage ensemble remains coherently coupled to the driven ensemble, which is positioned at the antinodes, through the coherent part of the DDI, enabling highly efficient excitation transfer. 

We observed that excitations can be quickly and efficiently transferred to the storage ensemble by carefully optimizing the driving strength to balance the coherent and dissipative timescales. Our findings further indicate that when the driven ensemble vastly outnumbers the storage ensemble ($N_{1} \gg N_{2}$), the driven ensemble can be treated as an effective classical excitation source, giving rise to a uniquely correlation-free inter-ensemble regime. To elucidate the excitation transfer in this regime, we developed a mean-field effective theoretical model, revealing the optimal transfer conditions and the associated scaling law with respect to the driven ensemble size $N_{1}$. Under the optimal condition, the storage ensemble can approach full inversion as the driven ensemble is scaled up, demonstrating a collective enhancement of the excitation-transfer process.

Building on this scaling mechanism, we proposed a three-stage storage-and-release protocol that integrates coherent excitation transfer, subradiant protection, and superradiant emission. Excitations are first efficiently routed into the subradiant storage ensemble, perfectly preserved via a frequency quench, and subsequently released on demand. Crucially, during the release stage, the macroscopic driven ensemble acts as a superradiant amplifier for the released photons, resulting in an enhanced peak emission rate. These results demonstrate that semi-infinite waveguide systems offer an ideal platform for rigorously engineering coherent and dissipative collective interactions. Our findings provide valuable insights and practical guidelines for developing scalable quantum memories, quantum batteries, and programmable nonequilibrium dynamics in waveguide QED architectures.

\section{Acknowledgments}
We are grateful to Kai-Min Hsieh for helpful discussions on the experimental realization. And we acknowledge support from the National Science and Technology Council (NSTC), Taiwan, under
the Grants No. 112-2112-M-001-079-MY3, NSTC-115-2112-M-001-035-MY3 and No. NSTC-115-2119-M-001-009, and from Academia Sinica under Grant AS-CDA-113-M04. We are also grateful for support from TG 1.2 of NCTS at NTU. GDL acknowledges support from Grant No. NSTC-113-2112-M-002 -025 and No. NSTC-112-2112-M-002 -001.

\begin{widetext}
\appendix
\section{The master equation}

The Hamiltonian describing the system shown in Fig.~\ref{config}(a) is $\hat{H} = \hat{H}_\text{S} + \hat{H}_\text{B} + \hat{H}_\text{int}$ \cite{lin2019scalable}, where (assuming $\hbar = 1$)
\begin{equation}
    \begin{cases}
        \hat{H}_\text{S} = \sum_i \omega_0 \hat{\sigma}^\dagger_i\hat{\sigma}_i, \\
        \hat{H}_\text{B} = \int^\infty_0 \text{d}\omega \omega \hat{a}^\dagger_{\omega} \hat{a}_{\omega}, \\
        \hat{H}_\text{int} = i \sum_i \int^\infty_0 \text{d} \omega g_i(\omega) \cos(k x_i) \hat{a}_{\omega} \hat{\sigma}^\dagger_i + \text{h.c.}.
    \end{cases}
\end{equation}
In the expression above, $\hat{a}_{\omega, s}$ is the bosonic annihilation operator of the waveguide mode with frequency $\omega$ propagating in $s = \pm$ direction, satisfying the commutation relation $[ \hat{a}_\omega, \hat{a}^\dagger_{\omega'}] = \delta(\omega - \omega')$. Due to the presence of the anti-node mirror at $x = 0$, the mode function in the interaction Hamiltonian is proportional to $\cos(kx)$. One may derive this by considering an open-ended regular 1D waveguide, then add image emitters with respect to the mirror. To derive Eq.~(\ref{ME_1}), we start from the Heisenberg equation of motion for the field operator:
\begin{equation}
    \dot{\hat{a}}_{\omega}(t) = -i\omega \hat{a}_\omega(t) + \sum_i g_i(\omega)\cos(kx_i)\hat{\sigma}_i(t).
\end{equation}
The solution to this equation is given by
\begin{equation}
    \hat{a}_\omega(t) = \hat{a}_\omega(0)e^{-i\omega t} + \sum_i g_i(\omega) \cos(kx_i) \int^t_0 dt' \hat{\sigma}_i(t')e^{-i\omega(t-t')}.
\end{equation}
Now we apply the Born-Markov approximation (extending the memory integral to $t \rightarrow \infty$ and evaluating $\hat{\sigma}_i(t')$ at time $t$) and plug $\hat{a}_\omega(t)$ into to the equation of motion of the full system density matrix $\tilde{\rho}_\text{tot}$ in the rotating frame with respect to $\hat{H}_\text{S} + \hat{H}_\text{B}$:
\begin{equation}
    \frac{\text{d}}{\text{d}t}\tilde{\rho}_\text{tot} = -i \left[ \tilde{H}_\text{int}(t), \tilde{\rho}_\text{tot} \right].
\end{equation}
By tracing out the field degrees of freedom assuming that the waveguide field is in a coherent state (input field), we come to the master equation presented in Eq.~(\ref{ME_1}).

\section{Analytical solution for $N_2 = 1$ excitation release}
\label{N_2=1}
Suppose that the system is initially prepared in the state
\begin{equation}
    \ket{\psi_0}
    =
    \ket{N_1/2,-N_1/2}_1\otimes\ket{e}_2,
\end{equation}
where $\ket{N_1/2,-N_1/2}_1$ is the lowest Dicke state of driven ensemble with total angular momentum $J=N_1/2$, and the emitter in ensemble 2 is initially excited. The subsequent dynamics is confined to the subspace spanned by
\begin{equation}
    \begin{cases}
        \ket{1}
        =
        \ket{N_1/2,-N_1/2}_1\otimes\ket{g}_2,
        \\[4pt]
        \ket{2}
        =
        \ket{N_1/2,-N_1/2+1}_1\otimes\ket{g}_2,
        \\[4pt]
        \ket{3}
        =
        \ket{N_1/2,-N_1/2}_1\otimes\ket{e}_2 .
    \end{cases}
\end{equation}
In this basis, the coherent coupling term can be written as
\begin{equation}
    g\sqrt{N_1}\hat{S}^{\dagger}_1\hat{\sigma}_2
    =
    gA_{-N_1/2}\ket{2}\bra{3},
\end{equation}
where $A_m=\sqrt{J(J+1)-m(m+1)}$, so that $A_{-N_1/2}=\sqrt{N_1}$. The collective decay operator of ensemble 1 becomes
\begin{equation}
    \sqrt{N_1}\hat{S}_1
    =
    A_{-N_1/2}\ket{1}\bra{2},
\end{equation}
and the initial state is therefore $\ket{\psi_0}=\ket{3}$. The master equation can then be reduced to the following equations of motion for the density-matrix elements:
\begin{equation}
    \frac{\text{d}}{\text{d}t}
    \begin{pmatrix}
        A(t) \\
        B(t) \\
        C(t)
    \end{pmatrix}
    =
    \begin{pmatrix}
        -\gamma N_1 & 2g\sqrt{N_1} & 0 \\
        -2g\sqrt{N_1} & -\gamma N_1 & \gamma N_1 \\
        0 & \gamma N_1 & -\gamma N_1
    \end{pmatrix}
    \begin{pmatrix}
        A(t) \\
        B(t) \\
        C(t)
    \end{pmatrix},
\end{equation}
where $A(t) = i\left(\rho_{23}-\rho_{32}\right)$, $B(t) = \left(\rho_{33}-\rho_{22}\right)$ and $C(t) = \left(\rho_{33}+\rho_{22}\right)$.
By solving these equations, and using $\rho_{22} = [C(t) - B(t)]/2$ and $g = \gamma$, we have
\begin{equation}
\rho_{22}(t)
=
    \begin{cases}
        \dfrac{2}{N_1-4}
        e^{-N_1\gamma t}
        \left[
        \cosh\!\left(
        \gamma\sqrt{N_1(N_1-4)}\,t
        \right)-1
        \right],
        & N_1>4,
        \\[10pt]
        4\gamma^2 t^2 e^{-4\gamma t},
        & N_1=4,
        \\[10pt]
        \dfrac{2}{N_1-4}
        e^{-N_1\gamma t}
        \left[
        \cos\!\left(
        \gamma\sqrt{N_1(4-N_1)}\,t
        \right)-1
        \right],
        & N_1<4 .
    \end{cases}
\end{equation}
Using the definition of the total emission rate, $\gamma_{\mathrm{tot}} = 2\gamma N_1 \langle \hat{S}^{\dagger}_1\hat{S}_1\rangle = 2N_1\gamma\rho_{22}$, we obtain the peak emission rate for $N_1>4$ as
\begin{equation}
    \gamma_{\mathrm{tot},\max}
    =
    2N_1\gamma
    \exp\!\left[
    -\sqrt{\frac{N_1}{N_1-4}}
    \ln\!\left(
    \frac{\sqrt{N_1}+\sqrt{N_1-4}}
         {\sqrt{N_1}-\sqrt{N_1-4}}
    \right)
    \right].
\end{equation}
In the large-$N_1$ limit, this expression approaches
\[
    \gamma_{\mathrm{tot},\max}
    \rightarrow
    2\gamma ,
    \qquad
    N_1\gg 1 .
\]

\section{Non-radiative decay and pure dephasing}
\label{nr_phi_deacy}

Here we model the nonradiative decay and dephasing by adding the corresponding dissipator to Eq.~(\ref{ME_1}) \cite{lin2019scalable}:
\begin{equation}
    \begin{cases}
        \mathcal{L}^\text{nr}[\hat{\rho}] = \sum_i 2\hat{\sigma}_i \hat{\rho} \hat{\sigma}^\dagger_i - \left\{\hat{\sigma}^\dagger_i\hat{\sigma}_i, \hat{\rho} \right\}, \\
        \mathcal{L}^{\phi}[\hat{\rho}] = \sum_i 2\hat{\sigma}^\dagger_i \hat{\sigma}_i \rho \hat{\sigma}^\dagger_i \hat{\sigma}_i - \left\{ \hat{\sigma}^\dagger_i \hat{\sigma}_i, \hat{\rho} \right\},
    \end{cases}
    \label{nr_phi_dissipator}
\end{equation}
with decay rate of $\gamma_\text{nr} = \epsilon \gamma$ and $\gamma_\phi = \epsilon\gamma$. Their effects are shown in Fig.~\ref{nr_decay} and Fig.~\ref{dephasing}, respectively. We simulate time evolution for $\langle n_2 \rangle$ and $\gamma_\text{tot}/\gamma$ with system size of $(N_1, N_2) = (7, 1)$. The release time is set to be $\gamma t = 5$ in all the diagrams. We extract the optimal excitation transfer $\langle n_2 \rangle_\tau$ and peak emission rate $\gamma_\text{tot, max}$ for different order of magnitude of $\epsilon$.
 
Although excitations in storage ensemble do not decay via waveguide mode, they're still susceptible to individual nonradiative decay. From the time evolution presented in Fig.~\ref{nr_decay}(a), we can see that the excitation decay away from stroage ensemble significantly during the storage stage for the case of $\epsilon = 0.05$, and smaller nonradiative decay rates do not largely disturb the performance of our scheme in finite time. Fig.~\ref{nr_decay}(b) and (c) shows that for $\epsilon \sim 10^{-2.5}$, the result of the simulation is similar to the case without nonradiative decay.

Pure dephasing has a much weaker effect on the performance of our scheme than nonradiative decay. As shown in Fig.~\ref{dephasing}(a), for the same order of $\epsilon$, the storage-and-release dynamics is only weakly disturbed by pure dephasing compared with the case of nonradiative loss. This difference arises because the storage mechanism relies on destructive interference between the field emitted by each emitter and that emitted by its own mirror image. In contrast to schemes based on antisymmetric dark states formed by two distinct emitters \cite{holzinger2022control, chen2025excitation, mirhosseini2019cavity, moreno2022efficient}, which are more vulnerable to decoherence between different emitters, the subradiance in our system is protected by the emitter--mirror interference and is therefore less sensitive to pure dephasing during the storage stage. Figures~\ref{dephasing}(b) and \ref{dephasing}(c) further show that pure dephasing noticeably reduces the performance only when the dephasing rate becomes large enough to induce significant decoherence during the driving and release stage.

\begin{figure}[h]
    \centering
    \includegraphics[width=0.8\linewidth]{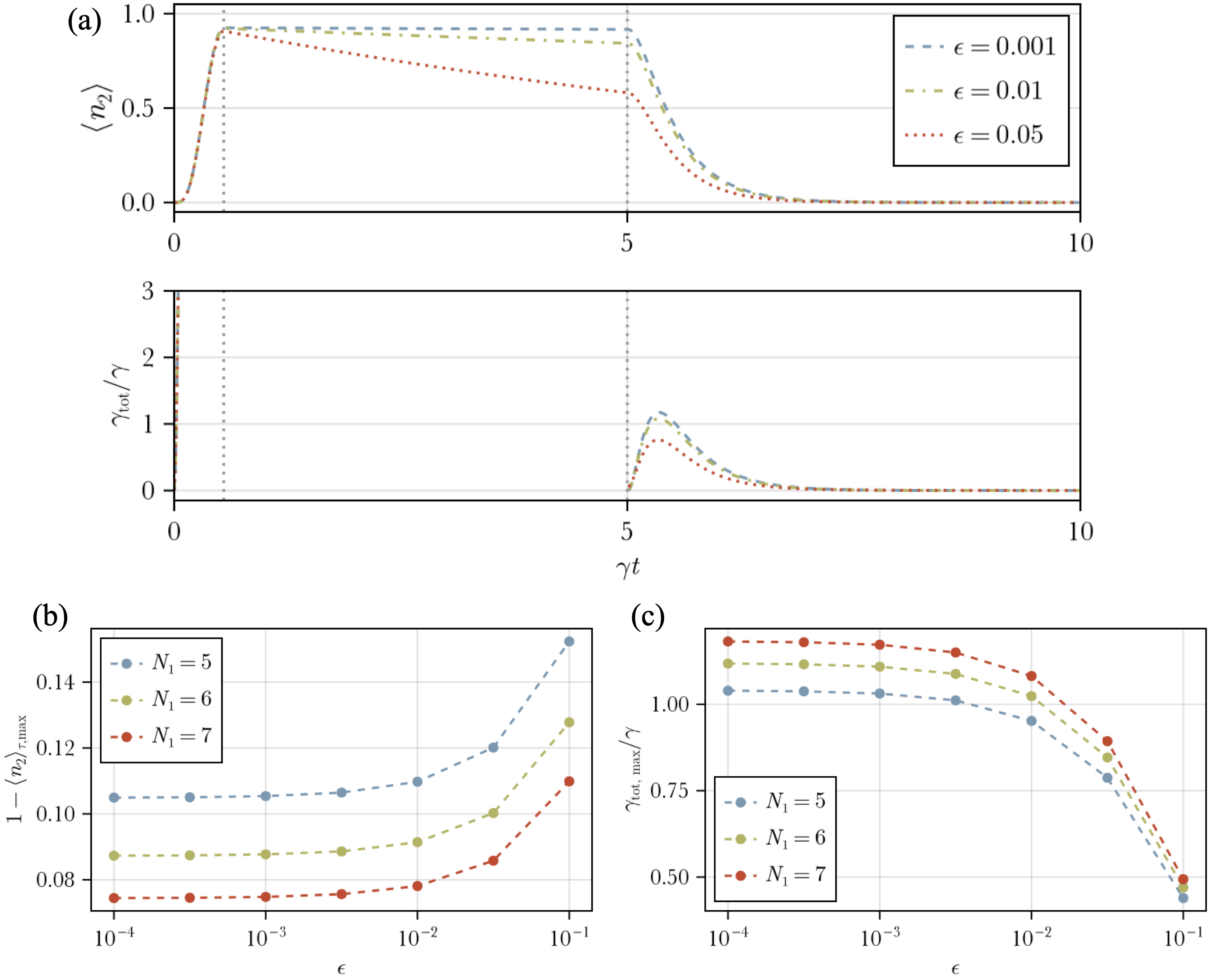}
    \caption{\textbf{The effect of nonradiative decay} (a) Time evolution of the excitation numbers and the emission rate, with system size of $(N_1, N_2) = (7, 1)$. The diagram shows the three stage process of the storage and release scheme including the effect of nonradiative decay with $\epsilon = 0.001, 0.01, 0.05$. (b, c) The mean excitation number at the start of the storage state $\langle n_2 \rangle_{\tau, \text{max}}$, and the peak emission rate $\gamma_\text{tot, max}/\gamma$, during the release stage for different values of $\epsilon$ for $N_1 = 6, 7, 8$. The release time for all the simulation is set to $\gamma t = 5$.}
    \label{nr_decay}
\end{figure}

\begin{figure}[H]
    \centering
    \includegraphics[width=0.8\linewidth]{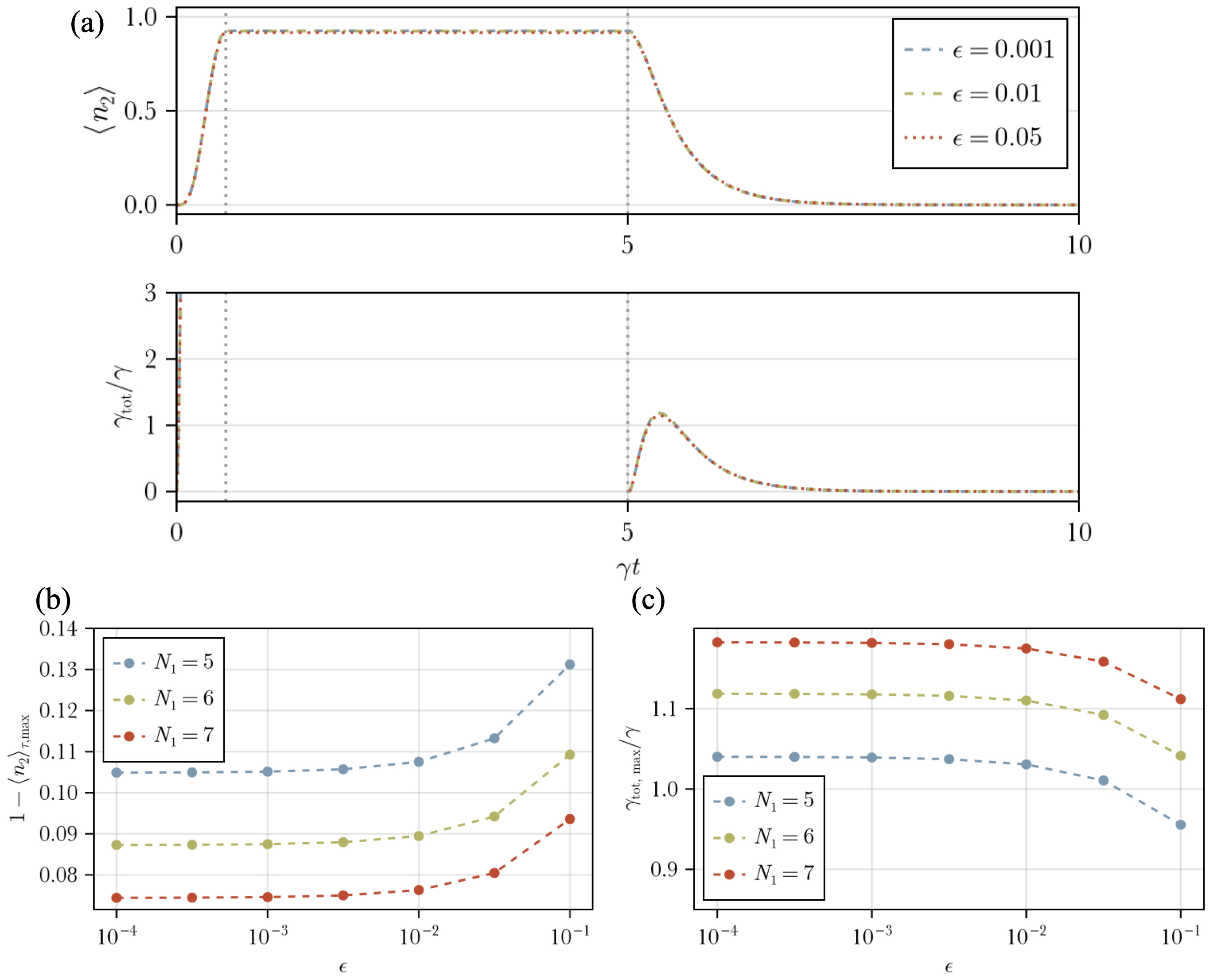}
    \caption{\textbf{The effect of pure dephasing} (a) Time evolution of the excitation numbers and the emission rate, with system size of $(N_1, N_2) = (7, 1)$. The diagram shows the three stage process of the storage and release scheme including the effect of pure dephasing with $\epsilon = 0.001, 0.01, 0.05$. (b, c) The mean excitation number at the start of the storage state $\langle n_2 \rangle_{\tau, \text{max}}$, and the peak emission rate $\gamma_\text{tot, max}/\gamma$, during the release stage for different values of $\epsilon$ for $N_1 = 5, 6, 7$. The release time for all the simulation is set to $\gamma t = 5$.}
    \label{dephasing}
\end{figure}
\end{widetext}

\bibliographystyle{unsrt}
\bibliography{ref}

\end{document}